# The walking behaviour of pedestrian social groups and its impact on crowd dynamics

Mehdi Moussaïd[1, 2, 3, *], Niriaska Perozo[1, 4], Simon Garnier[1], Dirk Helbing[2, 5], and Guy Theraulaz[1, 3]

[1] Université de Toulouse Paul Sabatier, Centre de Recherches sur la Cognition Animale, Toulouse, France.

[2] ETH Zurich, Swiss Federal Institute of Technology, Chair of Sociology, Zurich, Switzerland.

[3] CNRS, Centre de Recherches sur la Cognition Animale, Toulouse, France.

[4] Unidad de Investigación en Inteligencia Artificial, Universidad Centroccidental Lisandro Alvarado, Barquisimeto, Venezuela

[5] Santa Fe Institute, Santa Fe, USA

[*] Corresponding author: moussaid@cict.fr

**Abstract**

Human crowd motion is mainly driven by self-organized processes based on local interactions among pedestrians. While most studies of crowd behavior consider only interactions among isolated individuals, it turns out that up to 70% of people in a crowd are actually moving in groups, such as friends, couples, or families walking together. These groups constitute medium-scale aggregated structures and their impact on crowd dynamics is still largely unknown.

In this work, we analyze the motion of approximately 1500 pedestrian groups under natural condition, and show that social interactions among group members generate typical group walking patterns that influence crowd dynamics. At low density, group members tend to walk side by side, forming a line perpendicular to the walking direction. As the density increases, however, the linear walking formation is bent forward, turning it into a V-like pattern. These spatial patterns can be well described by a model based on social communication between group members. We show that the V-like walking pattern facilitates social interactions within the group, but reduces the flow because of its "non-aerodynamic" shape. Therefore, when crowd density increases, the group organization results from a trade-off between walking faster and facilitating social exchange.

These insights demonstrate that crowd dynamics is not only determined by physical constraints induced by other pedestrians and the environment, but also significantly by communicative, social interactions among individuals.

## *Introduction*

The study of human crowd dynamics has recently found great interest in many research fields [1,2,3,4,5]. In order to develop reliable prediction models for the design of urban infrastructures, traffic management or crowd safety during mass events or evacuation processes, it is necessary to understand the local interaction laws underlying collective crowd dynamics.

While a lot is known about the 'physics' of crowd motion, such as the organization emerging around bottlenecks [6,7], the segregation of opposite flows in pedestrian counterstreams [8,9,10], or the turbulent movement in extremely dense crowds [11,12], it is surprising that social interactions among pedestrians in crowd have been largely neglected. Indeed, the great majority of existing studies investigated a crowd as a collection of isolated individuals, each having an own desired speed and direction of motion, see e.g. Refs. [9,10,13,14]. In practice, however, it turns out that the majority of pedestrians actually do not walk alone, but in groups [15,16,17]. As we will show in this article, up to 70% of observed pedestrians in a commercial street are walking in group. Early observations have shown that groups composed of two to four members are the most frequent, while groups of size five and larger are rare. In addition, group sizes are distributed according to a Poisson distribution [17].

To our knowledge, however, the characteristics of the motion of pedestrian groups have not been empirically studied so far. It is basically unknown how moving group members interact with each other, with other pedestrians and with other groups. It also needs to be studied how such groups organize in space and how these spatial patterns affect the crowd dynamics. This is expected to be important for the planning of pedestrian facilities, mass events and evacuation concepts.

We note that the term 'group' is used here in its sociological sense [18], that is, not only referring to several proximate pedestrians that happen to walk close to each other, but to individuals who have social ties and *intentionally* walk together, such as friends or family members. In particular, the duration of the interaction and the communicative setting distinguish from an occasional agglomerate.

In this work, we analyze the organization of pedestrian social groups and their impact on the complex dynamics of crowd behavior. For this, we collected empirical data of the motion of pedestrian group by means of video recordings of public areas. Observations were made under low and moderate density conditions, called population A and B, respectively. We analyzed the behaviour of $N_A$=260 groups in population A and $N_B$=1093 groups in population B composed of two to four members (see Material & Methods). Relying on our observations, we developed an individual-based model of pedestrian behaviour. The model describes how an individual interacts with other group members and with outgroup pedestrians. By means of numerical simulations, we show that the model predicts the emergence of the empirically observed collective walking patterns well, and that pedestrian groups constitute a crucial component of the organization of human crowds.

## *Results*

### *Empirical observations*

According to our empirical analysis, the proportion of pedestrians belonging to a group is 55% in population A and 70% in population B, i.e. higher than the proportion of pedestrians walking alone. As shown in **figure 1**, the size of pedestrian groups in population A follows a zero-truncated Poisson distribution (p=0.06; on the basis of $\chi^2$-test), in agreement with previous observations [15,19]. In population B, the same tendency is observed, but the proportion of single pedestrians is lower than a Poisson distribution would predict, while the proportion of groups of size 2 is greater than expected (p<0.01). This difference between populations A and B is probably related to the environments in which the observations were made: While population A was observed during the afternoon of a working day, population B was observed on a Saturday in a popular commercial walkway, where one expects a higher tendency for people to have a leisure walk with friends. Effects of the social environment have also been observed in the past [15,19], namely the higher frequency of groups in leisure areas such as shopping centres or public beaches. Past studies have suggested that the observed size distribution could be explained by assuming that individuals would independently join and leave a group with a typical probability per unit of time, which implies that the rate of losing a member is proportional to the group size. According to analytical calculations, this mechanism can generate the observed distributions [15]. Next, we have measured the average walking speed of observed pedestrians **(figure 2)**. The speed of pedestrians is clearly dependent on the density level. At low density (population A), people walk faster than at higher density (population B). This is in agreement with previous empirical and theoretical studies of pedestrian traffic [20,21,22]. A new observation is that, in addition, pedestrian walking speeds decrease linearly with growing group size. Remarkably, the density level does not significantly affect the slope of the group-size-related speed decrease (ANCOVA, p=0.19, with *y=-0.04x+1.26* in population A and *y=-0.08x+1.24* in population B).

We then investigated the spatial organisation of walking pedestrian groups to find out whether there are any specific patterns of spatial group organization, and how such patterns may change with increasing density (see figure S1 of the supporting information). For this, we measured the average angle $\alpha_{ij}$ and distance $d_{ij}$ between pedestrians *i* and *j,* where *i* and *j* belong to the same group and *j* is *i*'s closest neighbour on the right-hand side*,* as sketched in **figure 3**. Numerical measurements for each group size and density level are provided in **table 1**. On the basis of the average angle and distance values for all pairs of pedestrian (*i , j)*, it is possible to reconstruct and visualize the observed patterns of spatial organization, as shown in **figure 4**.

At low density (population A), we observed that group members walked in a horizontal formation, where each pedestrian had his/her partners on the sides, at an angle of ±90° to the walking direction. A series of student t-tests revealed that the angle $\alpha_{ij}$ was not different from 90° for groups of size two ($p_{12}$>0.5), three ($p_{12}$=0.14; $p_{23}$>0.5), and four ($p_{12}$=0.13; $p_{23}$>0.5; $p_{34}$=0.47). This configuration facilitates social interactions within the groups because each member can easily communicate with his partners without turning the back to any of them.

At higher density levels (population B), the available space around the group is reduced. Group members can no longer maintain the same linear organization without interfering with out-group pedestrians. As shown in **table 1**, the average distances between group members was, in fact, reduced. Moreover, the configuration of the group changed: In groups of size 3, we observed that the middle pedestrian ($p_2$) tended to stand back, while the pedestrians $p_1$ and $p_3$ got closer to each other. This generated a 'V'-like formation, where the angle $\alpha_{12}$ was greater than 90° (108°±3; a unilateral t-test supports the difference from 90° with a value of p>0.5) and angle $\alpha_{23}$ is lower than 90° (71°±2; p>0.5 by unilateral t-test). In the same way, for groups of size 4, pedestrians $p_2$ and $p_3$ tend to move back, leading to a 'U'-like formation (a series of t-tests confirms that $\alpha_{12}$ is greater than 90° with a value of p>0.5, $\alpha_{23}$ is not different from 90° with p=0.21, and $\alpha_{34}$ is smaller than 90° with a value of p>0.5). Therefore, the horizontal walking formation observed at low density is bent when the density level increases, allowing the group to occupy a smaller area. However, it is surprising that the bending is *forward* in walking direction, not *backward* as expected for a flexible structure moving against an opposite flow. This suggests that this structure is *actively* created and maintained in order to support certain functions (e.g. better communication).

*Mathematical model*

To better understand the above empirical results, we extend an existing model of pedestrian behavior to include social interactions among people walking in groups. For this, we rely on the experimental specification of the social force model, that has been experimentally calibrated and validated in a previous work [9]. The basic modelling concept suggests that the motion of a pedestrian *i* can be described by the combination of a driving force $\vec{f}_i^{\,0}$ that reflects a pedestrian's motivation to move in a given direction at a certain desired speed, a repulsive force $\vec{f}_{ij}$ describing the effects of interactions with other isolated pedestrians *j*, and $\vec{f}_i^{\,wall}$ reflecting the repulsive effects of boundaries such as walls or obstacles in streets (see Material & Methods for the mathematical specification of these interactions forces).

In this section, we formulate a new interaction term $\vec{f}_i^{\,group}$ describing the response of pedestrian *i* to other group members. Therefore, the complete equation of motion reads

$$\frac{d\vec{v}_i}{dt} = \vec{f}_i^{\,0} + \vec{f}_i^{\,wall} + \sum_j \vec{f}_{ij} + \vec{f}_i^{\,group}.$$

We postulate that the observed patterns of group organization result from the desire of their respective members to communicate with each other. Therefore, individuals continuously adjust their position to facilitate verbal exchange, while trying to avoid collisions with in-group members and out-group pedestrians. In particular, it has been shown that the gaze direction and eye contact are essential features of group communication, as it helps to get a feedback about the other persons' reactions [23,24,25]. Accordingly, we introduce a vision field as an important component of our pedestrian simulation model.

In a group of size *N*, we define a gazing direction vector $\vec{H}_i$ for each of its members *i*. The angle of vision of pedestrian *i* is $\phi$ degrees to the left and to the right of the gazing direction. In addition, we define the point **c<sub>i</sub>** as the centre of mass of all other group members walking with pedestrian *i* (**figure 5**).

In our computer simulations, group members turn their gazing direction to see their partners. To do so, the gazing direction vector $\vec{H}_i$ is rotated by an angle $\alpha_i$, so that point $\mathbf{c_i}$ is included in the vision field of pedestrian *i* (as sketched in **figure 5**).

However, the greater $\alpha_i$, the less comfortable is the turning for walking. Therefore, we assume that pedestrian *i* adjusts its position to reduce the head rotation $\alpha_i$. This is modeled by the acceleration term

$$\vec{f}_i^{\,vis} = -\beta_1 \alpha_i \vec{V}_i,$$

where $\beta_1$ is a model parameter describing the strength of the social interactions between group members, and $\vec{V}_i$ is the velocity vector of pedestrian *i*. The related deceleration is assumed to be proportional to the head rotation $\alpha_i$. At the same time, pedestrian *i* keeps a certain distance to the group's center of mass. According to our observations, the average to the center of mass increases with group size. Therefore, we define a second acceleration term

$$\vec{f}_i^{\,att} = q_A \beta_2 \vec{U}_i,$$

where $\beta_2$ is the strength of the attraction effects and $\vec{U}_i$ is the unit vector pointing from pedestrian *i* to the center of mass. Furthermore $q_A=1$ if the distance between pedestrian *i* and the group's centre of mass exceeds a threshold value, otherwise $q_A=0$. According to the data collected under low density conditions, the threshold value can be approximated as $\frac{(N-1)}{2}$ meters.

Finally, we add a repulsion effect so that group members do not overlap each other, which is simply defined as

$$\vec{f}_i^{\,rep} = \sum_k q_R \beta_3 \vec{W}_{ik}.$$

Here, $\vec{W}_{ik}$ is the unit vector pointing from pedestrian *i* to the group member *k*, and $\beta_3$ is the repulsion strength. Moreover, $q_R=1$ if pedestrians *i* and *k* overlap each other (when the distance $d_{ik}$ is smaller than a threshold value $d_o$, that is one body diameter plus some safety distance), otherwise $q_R=0$.

In summary, the social interaction term $\vec{f}_i^{\,group}$ is defined as:

$$\vec{f}_i^{\,group} = \vec{f}_i^{\,vis} + \vec{f}_i^{\,att} + \vec{f}_i^{\,rep}.$$

*Simulation results*

Computer simulations of the above model were performed in a way reflecting the empirical conditions of populations A and B (see Material & Methods). As for the observed data, we measured the average angle and distance between each pair of pedestrians, and studied the related pattern of organization. Simulated groups form collective walking patterns that match the empirical ones very well (see **figure 4**). In particular, a series of Student t-tests reveals no significant difference between the observed angle distributions and the predicted ones (see the table S1 in Supporting information). The spatial pattern of the group is mainly influenced by parameter $\beta_1$, representing the strength of the social interactions between group members (**figure 6**). When setting $\beta_1=0$, group members only try to stick together with no communication rule, and tend to form an "aerodynamic" *inverse* V-like shape. In contrast, for the realistic value $\beta_1=4$, groups form the observed forwardly directed V-like pattern, which, however, affects the overall walking speed of the crowd.

In accordance with empirical results, the model predicts a linear decrease of the walking speeds with increasing group size, with a similar slope for both density levels. An ANCOVA test delivers a p-value of 0.071 thereby accepting the hypothesis that the slopes are not different, with *y=-0.05x+1.3* at low density and *y=-0.07x+1.2* at moderate density.

## *Discussion*

When studying crowd dynamics, the majority of previous publications have neglected the influence of pedestrians groups. Despite past observations revealing the existence of groups in pedestrian crowds, nothing was known about the spatial organization of moving groups and their impact on the overall crowd dynamics. Combining empirical observations with a properly extended interaction model, we have shown how social interactions among group members generate a typical group organization.

Our empirical observations reveal that much of pedestrian traffic is actually made up of groups. In our data, only one third of observed pedestrians were walking alone. Furthermore, it turns out that pedestrian groups have an important impact on the overall traffic efficiency. This underlines the necessity to consider groups in futur studies of pedestrian dynamics.

We found that typical group walking patterns emerge from local interactions among group members. At low density, group members tend to walk side-by-side, forming a line perpendicular to the walking direction, thereby occupying a large area in the street. Hence, when the local density level increases, the group needs to adapt to the reduced availability of space. This is done by the formation of 'V'-like or 'U'-like walking patterns in groups with three or four members, respectively. As shown by numerical simulations, these configurations are emergent patterns resulting from the tendency of each pedestrian to find a comfortable walking position supporting communication with the other group members.

However, the walking efficiency is considerably affected by the fact that 'V'-like and 'U'-like configurations are *convex* shapes, which do not have optimal 'aerodynamic' features. Indeed, a *concave* shape, such as an inverse 'V' shape, would be advantageous since it would support the movement against a flow of people (as the flight formation of migrating birds such as geese or ducks reduces the aerodynamic friction [26,27]).

Additional computer simulations show that the model parameter $\beta_1$ representing the strength of social interactions among group members is essential to capture the dynamics of the system (see **figure 6a)**. When $\beta_1$ is set to 0 (i.e. when group members would only try to stick together with no communication rule), an inverse 'V'-like configuration is generated and the walking speed is close to a situation with isolated individuals only (compare the dashed and dark grey curves in **figure 6**). In contrast for $\beta_1=4$, the value determined from our empirical results, the speed is reduced by an average of 17% (see light grey curve). Therefore, two conflicting tendencies are involved: to walk fast and efficiently at minimum 'friction' (generating an *inverse* 'V'-like configuration), and to have social interactions with group members (supporting a 'V'-like configuration). At very low density, both tendencies are compatible, as pedestrians can walk side by side at a speed close to the desired one. At moderate densities, however, it appears that the social interactions are given a greater importance, supporting a V and U-like configuration, as empirically observed. However, it could happen that, when the density reaches very high levels, the physical constraints would prevail over the social preferences, and group members would start walking one behind another, forming a 'river-like' following pattern, as reported by Helbing et al. [28].

One may also ask how groups with more than four members would organise. It is, in fact, unlikely that a group of ten people would still walk side by side. This would require that each group member wanted to communicate with all the others at the same time. Instead, it expected that large groups (such as tourists or hiking groups) would typically split up. The most plausible explanation for group splitting is that, when group members are too far away from each other to communicate, they only consider those in the immediate surrounding. Consequently, clusters of two to four people would emerge within the group. In our model, this could be implemented by specifying the focus point $c_i$ of pedestrian $i$ not as the centre of mass of *all* other group members, but only a *few* of them.

In addition, one may expect a leader effect in pedestrian groups. For example, it is known that the distribution of spoken contributions among group members is not equal during a conversation. It rather follows a Poisson distribution, where a few members speak most of the time, while the others listen [29,30]. Therefore, it is likely that pedestrians who talk more would end up in the middle of the group and the listeners would walk on the sides. In the same way, large groups would probably split up into subgroups around those who talk most. It will be interesting to test this hypothesis experimentally in the future.

In summary, social interactions are a crucial aspect of the organization of human crowds, which should to be taken into account in future studies of crowd behavior.

## *Material and Methods*

### *Ethics statement*

No ethics statement is required for this work. Video recordings of pedestrian crowds were made in public areas and the data were analyzed anonymously.

### *Empirical observations*

The data for population A were collected during spring 2006 in a public place in the city of Toulouse, France, while data for population B were collected during spring 2007 in a crowded commercial walkway on a Saturday afternoon. Observations were made with a digital camera (SONY DCR-TRV950E, 720x576 pixels) during two hours at a frequency of one frame per second and five frames per second for population A and B, respectively. Pedestrian positions were then manually tracked by means of a dedicated software developed in our team, and their coordinates were reconstructed after correction of the camera lens distortion. A total of 1098 and 3461 pedestrians were tracked in population A and B respectively. People belonging to the same group were identified with a series of criteria defined in previous studies on pedestrian groups [17]. In particular, group membership was identified by a clear social interactions among group members, such as talking, laughter, smiles or gesticulation. On average, populations A and B were characterized by global density levels of 0.03 and 0.25 peds/m$^2$, respectively. In both populations, the speed of each group was computed as the average speed of its group members. Groups which temporarily stopped their motion were detected according to the procedure described by Collins et al. [31] and not considered in the computation of the average walking speed and the spatial patterns (but included in the density measurement).

*Model and Simulation Design*

According to previous work, the motion of an isolated pedestrian $i$ can be well described by means of three different acceleration components [10]: (1) the acceleration behavior $\vec{f}_i^{\,0}$, reflecting the pedestrian's desire to move in a particular direction at a certain speed, (2) repulsive effects $\vec{f}_i^{\,wall}$ on the pedestrian due to boundaries, and (3) interaction effects $\vec{f}_{ij}$, reflecting the response of pedestrian $i$ to other pedestrians $j$.

The acceleration behavior $\vec{f}_i^{\,0}$ was experimentally measured in past studies [9] and can be well described by

$$\vec{f}_i^{\,0} = \frac{d\vec{v}_i}{dt} = \frac{v_i^0 \vec{e}_i^{\,0} - \vec{v}_i(t)}{\tau}.$$

This relationship reflects the adaptation of the current velocity $\vec{v}_i$ of pedestrian $i$ to a desired speed $v_i^0$ and a desired direction of motion $\vec{e}_i^{\,0}$ within a certain relaxation time $\tau$. The empirically determined parameter values are $v_i^0 = 1.3 m/s$ and $\tau = 0.5s$.

Interactions $\vec{f}_i^{\,wall}$ with the boundaries have been specified in agreement with previous findings [13], i.e. as an exponentially decaying function of the distance $d_w$ perpendicular to the boundary: $\vec{f}_i^{\,wall}(d_w) = ae^{-d_w/b}$. The parameters $a = 10$ and $b = 0.1$ reflect that the wall repulsion extends over 30cm.

Finally, the pedestrian interactions $\vec{f}_{ij}$ have been specified according to the experimental model described by Moussaïd et al. [9].

The model parameters given in the caption of figure 4 represent the calibration result of a systematic scan of the parameter space, during which group motion was simulated with parameter values from reasonable ranges, identifying the parameter combination that generated the best agreement with the empirical observations. The comparison with the empirical data was made on the basis of the average angle and distance values between pedestrians given in Table 1. In our computer simulations, pedestrians started with random positions and with a random specification of the walking direction parallel to the street. Members of a group started one meter away from each other, having the same desired walking direction. The desired speeds were normally distributed with mean value 1.3m/s and standard deviation 0.2 m/s, to reflect the natural variability of pedestrian behavior. The simulations were performed with periodic boundary conditions. In order to reflect the environment where the data were collected, the street dimension was set to 18x18 meters and 5x14 meters for the low-density scenario and the moderate-density scenario, respectively. The number $N_s$ of groups of size *s* was specified in such a way that the density level in the simulation was the same as the empirically observed one for population A and B, i.e. $N_1=2$, $N_2=1$, $N_3=1$, and $N_4=1$ at low density corresponding to population A, and $N_1=5$, $N_2=2$, $N_3=1$, and $N_4=1$ at moderate density corresponding to population B. Measurements were made after 10 seconds of simulation which was enough for the walking patterns to appear, and over a time period of 5 seconds. Simulation results were averaged over 1000 runs. The time step was set to $\delta t = 1/20 s$.

## Acknowledgements

We thank Christian Jost, Jacques Gautrais, Alexandre Campo, Jeanne Gouëllo, Mathieu Moreau and the members of the EMCC research group in Toulouse for inspiring discussions, as well as four anonymous reviewers for helpful comments on the manuscript.

## Figures

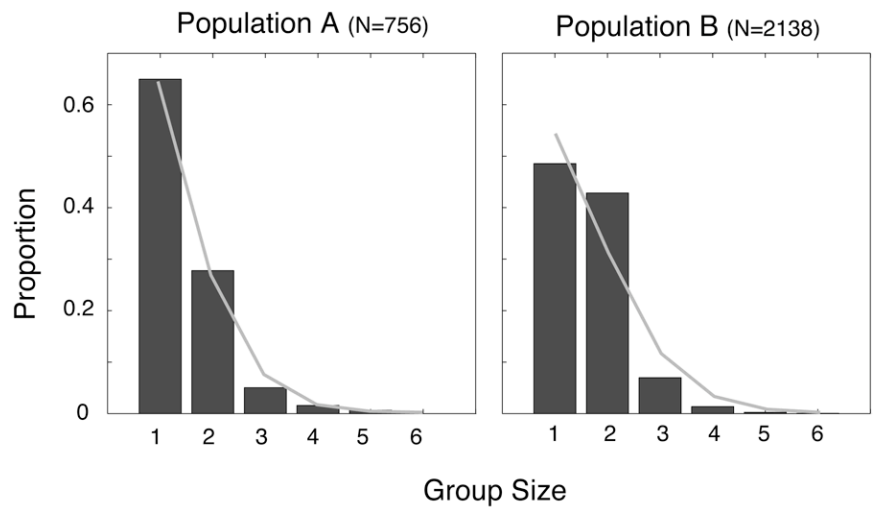

**Figure 1:** Observed group size distribution in populations A and B. The light grey curve indicates the zero-truncated Poisson fit ($N_i = e^{-\lambda} \frac{\lambda^i}{i!(1-e^{-\lambda})}$) with $\lambda = 0.83$ and $\lambda = 1.11$ for populations A and B, respectively.

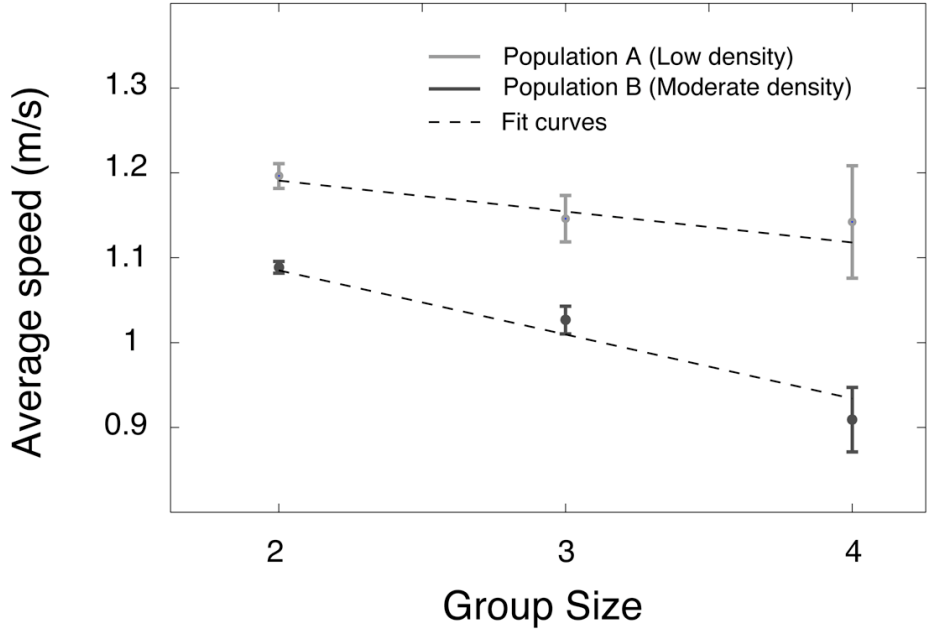

**Figure 2:** Effects of group size on walking speed. Average walking speed as a function of group size at low density (light grey) and moderate density (dark grey). Error bars indicate the standard error of the mean value. The fit curves are *y=-0.04x+1.26* for population A and *y=-0.08x+1.24* for population B.

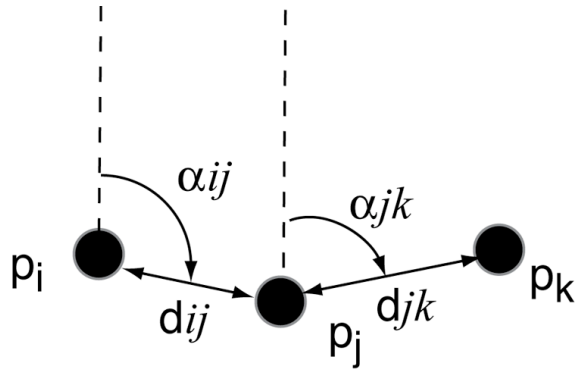

**Figure 3:** Illustration of the measurement method. We define $\alpha_{ij}$ and $d_{ij}$ as the angle and distance between pedestrians *i* and *j*, where *i* and *j* belong to the same group and *j* is *i*'s closest neighbour on the right-hand side.

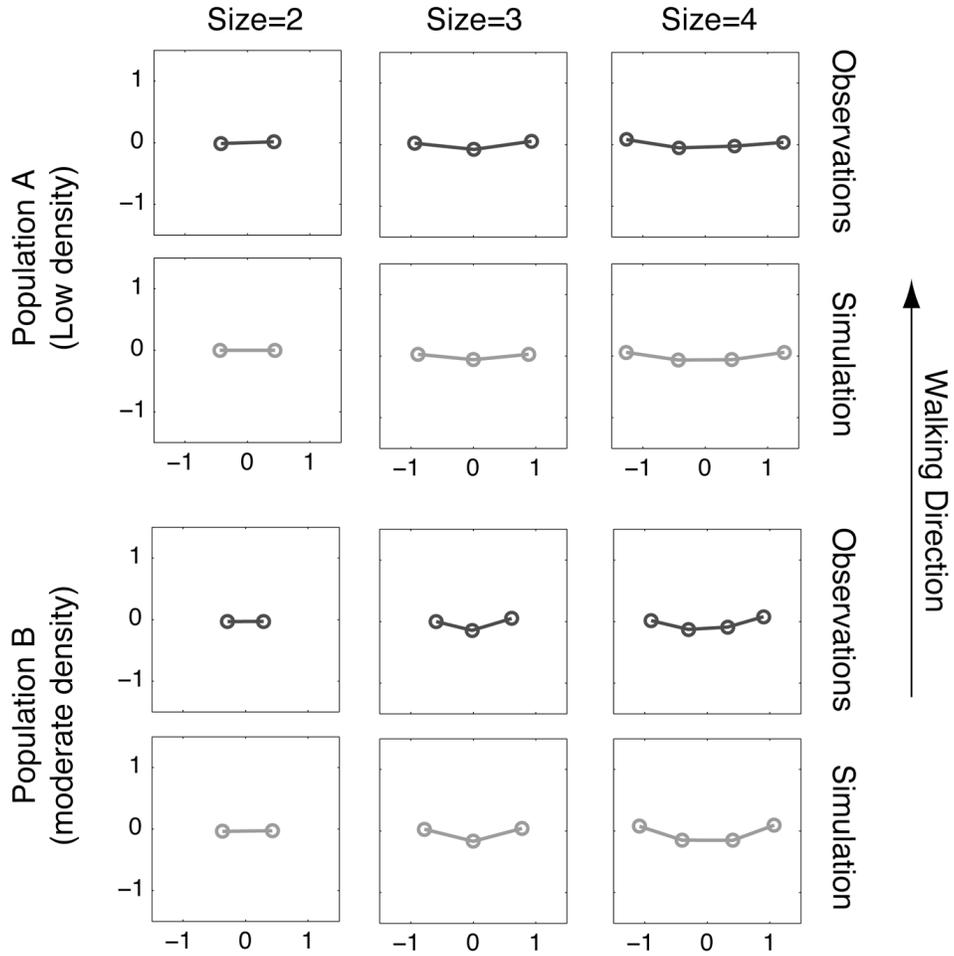

**Figure 4:** Average patterns of organization. The positions of pedestrians are reconstructed from the empirical angle and distance values provided in table 1 (dark grey), and from simulation results (light grey). The best fit parameters were obtained through a calibration process and amount to $\beta_1 = 4$; $\beta_2 = 3$; $\beta_3 = 1$; $d_o = 0.8$m; $\phi = 90°$.

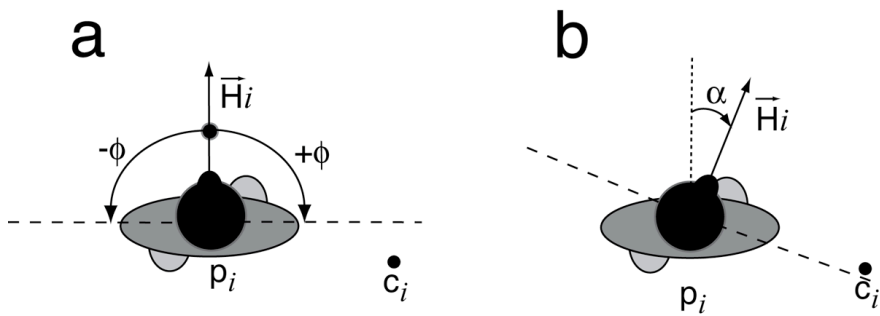

**Figure 5:** Illustration of the model variables. a) $\vec{H}_i$ is the gazing direction vector of pedestrian $i$. The dashed lines represent the borders of the visual field. b) Pedestrian $i$

rotates his head direction by an angle $\alpha$, so that the focus point $c_i$ is included in the vision field.

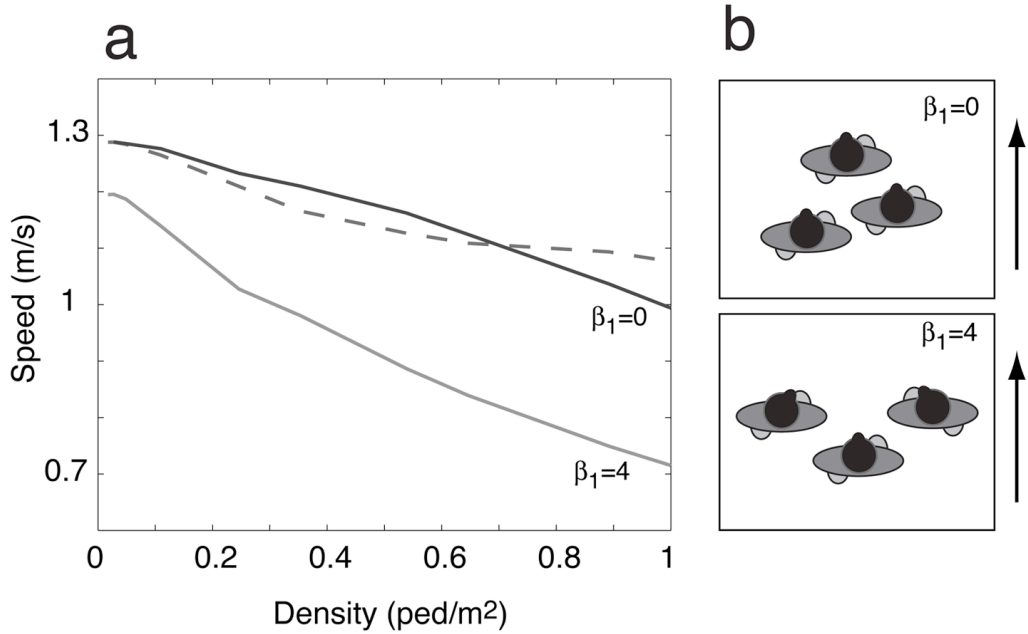

**Figure 6:** Simulation results for pedestrian groups with and without communication-enhancing interactions. (a) Speed-density curves showing the impact of group organization on traffic efficiency. For $\beta_1=0$, group members are attracted by the group's centre of mass only letting them stay together. This creates an *inverse* V-shaped configuration. For $\beta_1=4$, the value determined from our empirical observations, group members adapt their position to see the other group members, creating a V-shaped configuration. The dashed curve corresponds to simulations with isolated pedestrians only (no groups). (b) Illustration of typical group patterns for $\beta_1=0$ and $\beta_1=4$ at a density of 0.25 ped/m². The simulation parameters are the same as in figure 4.

**Table**

|  |  | Population A | | Population B | |
|---|---|---|---|---|---|
|  |  | $\alpha_{ij}$ (deg) | $d_{ij}$ (m) | $\alpha_{ij}$ (deg) | $d_{ij}$ (m) |
| Size=2 | $p_1p_2$ | 89.8 (±1.12) | 0.78 (±0.02) | 90.3 (±0.80) | 0.54 (±0.01) |
| Size=3 | $p_1p_2$ | 97.8 (±5.14) | 0.79 (±0.05) | 107.9 (±2.84) | 0.55 (±0.01) |
|  | $p_2p_3$ | 87.1 (±4.46) | 0.81 (±0.10) | 70.6 (±2.55) | 0.62 (±0.04) |
| Size=4 | $p_1p_2$ | 99.2 (±6.33) | 0.87 (±0.06) | 102.3 (±5.85) | 0.67 (±0.02) |
|  | $p_2p_3$ | 87.7 (±6.54) | 0.93 (±0.09) | 86.0 (±4.71) | 0.66 (±0.02) |
|  | $p_3p_4$ | 85.4 (±5.01) | 0.80 (±0.05) | 76.6 (±5.09) | 0.64 (±0.03) |

Average angle and distance values between group members for each group size and density level. Values between brackets indicate the standard error of the mean.